\documentclass[letterpaper, 10 pt, conference]{ieeeconf}  

\usepackage{graphicx}      

\usepackage{tikz}
\usepackage{amsmath}
\usepackage{amssymb}
\usepackage{amsfonts}

\newtheorem{theorem}{Theorem}

\newtheorem{proposition}[theorem]{Proposition}

\usepackage{algorithm,algorithmic}
\usepackage{epstopdf}

\usepackage{todonotes}
\usepackage{caption}
\usepackage{subfig}
\usepackage[maxfloats=200]{morefloats}
\usepackage{mathrsfs}

\usepackage{mathtools}

\usepackage{placeins}
\IEEEoverridecommandlockouts                              

\title{\LARGE \bf
Safe Output Feedback Improvement with Baselines
}

\author{Ruoqi Zhang$^{1}$, Per Mattsson$^{1}$ and Dave Zachariah$^{1}$ 
\thanks{*This work was supported by the \emph{Swedish Research Council (VR)} under contract 2023-04546.}
\thanks{$^{1}$Department of Information Technology, Uppsala University,
   SE-75105 Uppsala, SWEDEN (e-mail: ruoqi.zhang@it.uu.se, per.mattsson@it.uu.se, dave.zachariah@it.uu.se).}}%

\usepackage{xcolor}

\begin{document}

\makeatletter
\def\endthebibliography{%
  \def\@noitemerr{\@latex@warning{Empty `thebibliography' environment}}%
  \endlist
}
\makeatother

\maketitle
\thispagestyle{empty}
\pagestyle{empty}

\begin{abstract}
In data-driven control design, an important problem is to deal with uncertainty due to limited and noisy data. 
One way to do this is to use a min-max approach, which aims to minimize some design criteria for the worst-case scenario. 
However, a strategy based on this approach can lead to overly conservative controllers. 
To overcome this issue, we apply the idea of baseline regret,  
and it is seen that minimizing the baseline regret under model uncertainty can guarantee safe controller improvement with less conservatism and variance in the resulting controllers.
To exemplify the use of baseline controllers, we focus on the output feedback setting and propose a two-step control design method; first,
an uncertainty set is constructed by a data-driven system identification approach based on finite impulse response models; then a control design criterion based on model reference control is used.  
To solve the baseline regret optimization problem efficiently, we use a convex approximation of the criterion and apply the scenario approach in optimization. 
The numerical examples show that the inclusion of baseline regret indeed improves the performance and reduces the variance of the resulting controller.

\end{abstract}

\section{INTRODUCTION}
In recent years, there has been an increased interest in data-driven control design. A key issue in this setting is to achieve a controller that is robust to uncertainties due to limited and noisy data. A classical approach to handle this is to use a min-max criterion \cite{scokaert1998minmax, bemporad2003minmax}, which tries to find the controller that performs best in the worst-case scenario. In \cite{anna2019bayesian} it is shown how this idea can be used in traditional output feedback control. 
Through this stringent and pessimistic consideration, the min-max approach provides a performance guarantee, thereby yielding a safe controller.
Nevertheless, the resulting controller may be overly conservative \cite{mayne2000constrained, alessio2009survey}, leading to slow response and inefficient operation. 

However, in many real-world scenarios, an operational controller already exists. These controllers are often simple in design and implementation, for example, a proportional controller \cite{ruoqi-robust-acc}, and are designed by utilizing prior knowledge about the system from expert knowledge or insights derived from similar systems. In such cases, it makes sense to use this existing controller as a baseline, and then aim to find a controller that improves over the baseline. In this paper we will see that such a strategy can reduce the conservatism of a pure min-max approach substantially. 

The idea of using a baseline controller like this has been explored in the reinforcement learning literature. In \cite{ghavamzadeh2016safe_baseline, laroche2019safe} a safe policy improvement method is delineated by minimizing the negative regret with respect to the baseline policy over robust Markov Decision Processes \cite{nilim2005robust}. 
However, these methods do not directly translate to traditional output feedback controllers, since they assume full state information. In many real-world situations, the system is only partially observable implying that only outputs are accessible with observation noises. 
\begin{figure}
    \centering
    \includegraphics[width=0.9\linewidth]{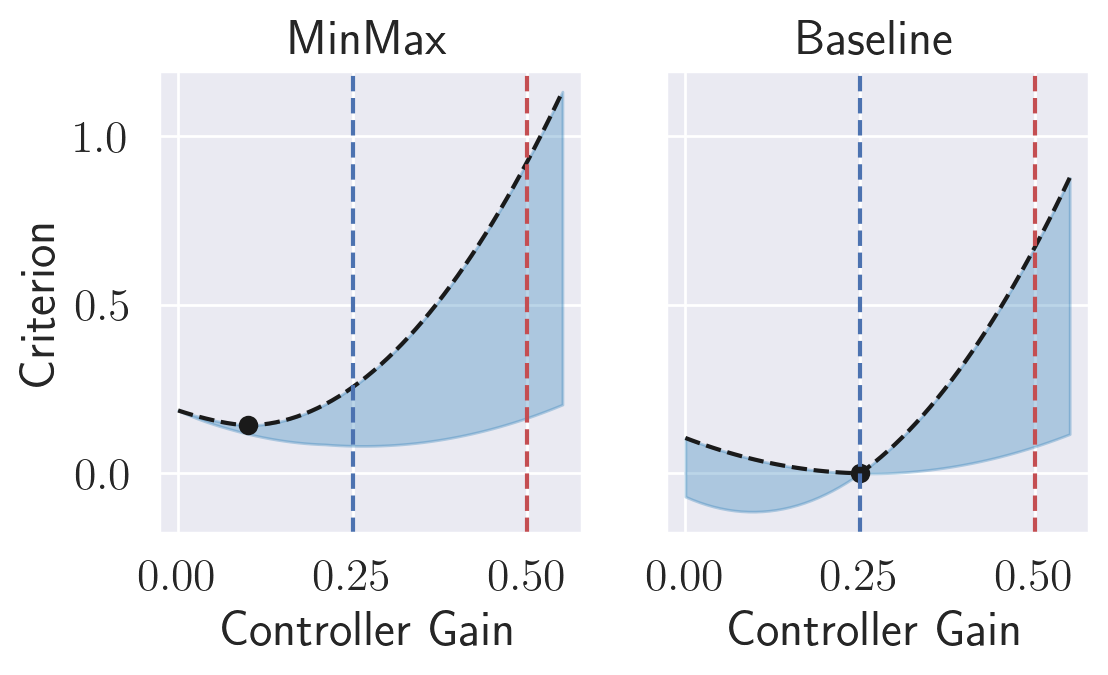}
    \caption{Controller gain versus performance criterion. Left: The \emph{black dashed line} shows the worst-case cost under model uncertainty represented by shaded area and the \emph{dot} indicates the min-max $K$, see \eqref{eq:min-max}. 
    The \emph{blue dashed line} shows the baseline $K_b$.
    The \emph{red dashed line} indicates the optimal K for the unknown system $G_{\circ}$. 
Right: The \emph{black dashed line} shows worst-case baseline regret and the dot indicates the proposed controller, see \eqref{eq:min-max:baseline}. With high uncertainty, the min-max strategy chooses a conservative controller with a low gain while our method selects the baseline controller.}
    \label{fig:1D-case-set}
\end{figure}
In this work we start from a setting similar to that of \cite{anna2019bayesian, ferizbegovic2021bayes}, where pure min-max criteria are considered. Such a criterion is illustrated in the left part of Figure~\ref{fig:1D-case-set}, where the aim is to adjust the gain in a P-controller. In cases where the uncertainty about the system is high, we can see that we get a controller gain that is much lower than the optimal gain (red dashed line), and thus a very conservative controller. 

The main contribution of this paper is to introduce a baseline controller (blue dashed line) into the criterion, resulting in the right part of Figure~\ref{fig:1D-case-set}. By considering robust improvements compared to the baseline instead of the pure min-max criterion, we can avoid being unnecessarily conservative. While this principle can be deployed in different settings, we illustrate it with a focus on output feedback controller design using a model reference criterion similar to \cite{anna2019bayesian, ferizbegovic2021bayes}. To find an uncertainty set in a data-driven manner we consider FIR systems of high order, to be able to capture general (stable) linear dynamics. Finally, to get a tractable convex optimization problem we utilize approximations akin to \cite{anna2019bayesian} in conjunction with the scenario approach \cite{calafiore2006scenario, campi2009scenario, campi2021scenario}. 
The effect of adding a baseline controller to the standard min-max criterion is illustrated in numerical examples, showing that the resulting controllers have less variance and are less conservative than a pure min-max controller. 

The remainder of the paper is organized as follows: In Section~\ref{sec::ps} the min-max and baseline regret optimization problems are defined and discussed. In Section~\ref{sec::sets} a system identification approach is used to find an uncertainty set in a data-driven manner. In Section~\ref{sec::convex} the approximations needed to get a convex optimization problem are presented, and in Section~\ref{sec::num} the methods are evaluated on two numerical examples. We conclude the paper in Section~\ref{sec::conclusions}.

\section{BASELINE REGRET}
\label{sec::ps}
Consider a stable discrete-time linear system 
\begin{align}
	y_t = G_{\circ}(q)u_t + v_t 
\end{align}
where $u_t$ is the input, $y_t$ is the output and $v_t$ is noise. $G_{\circ}(q)$ is an unknown transfer operator and $q^{-1}$ is the backward shift operator, $q^{-1} u_t = u_{t-1}$. 
The aim is to design a linear output feedback controller 
\[
u_t = C(q) (r_t - y_t)
\]
such that some performance criterion $J$ is minimized,
\begin{align}
    C \in \arg\min_{C\in \mathcal{C}} J(C,G), \nonumber
\end{align}
where $\mathcal{C}$ is a set of all feasible controllers, see for example~\cite{anna2019bayesian}. 
Since $G_{\circ}$ is unknown, we will use open-loop experimental data $(u_t, y_t)$, $t=1, \ldots, N$, to estimate an uncertainty set $\mathcal{G}$ that contains the true system with high probability. Given such a set, a common approach to be robust to the uncertainty involves the utilization of min-max optimization
\begin{align}
	\label{eq:min-max}
	C \in \arg \min_{C \in \mathcal{C}} \max_{G\in \mathcal{G}}J(C, G).
\end{align} 

The main drawback with this approach is that the obtained controller can be overly conservative if the uncertainty set $\mathcal{G}$ is large. 
To overcome this problem, we will introduce a baseline controller. 
That is, assume we have access to a controller $C_b(q) \in \mathcal{C}$ with acceptable, but suboptimal, behavior. The question then is if we can find a controller that improves the performance compared to the baseline controller for every potential system within $\mathcal{G}$. This leads us to a paradigm centered on minimizing the maximum baseline regret,
\begin{align}
	\label{eq:min-max:baseline}
	C \in \arg \min_{C \in \mathcal{C}} \max_{G\in \mathcal{G}}J(C, G) - J(C_b,G).
\end{align}

We call this safe controller improvement.
This strategy ensures that, across the entirety of $\mathcal{G}$, the regret relative to the baseline is non-positive, since the choice $C=C_b$ in the minimization makes the criterion in \eqref{eq:min-max:baseline} equal to zero. 
In particular, this means that if the true, but unknown, $G_o$ is in $\mathcal{G}$, then the controller given by \eqref{eq:min-max:baseline} is at least as good as the baseline controller for the true system.
\begin{proposition}
If $G_{\circ} \in \mathcal{G}$, and the solution to \eqref{eq:min-max:baseline} is $\hat{C}^*$, then {$J(\hat{C}^*, G_{\circ}) \leq J(C_b,G_{\circ})$.} 
\end{proposition}
\begin{proof}
If the true system $G_{\circ} \in \mathcal{G}$, it can be guaranteed that $J(\hat{C}^*,G_{\circ})-J(C_b, G_{\circ}) \leq \max_{G\in \mathcal{G}} J(\hat{C}^*, G) - J(C_b,G) \leq  J({C}_b, G) - J(C_b,G) = 0$. 
\end{proof}

Hence, if we can find an uncertainty set $\mathcal{G}$ that contains $G_o$ and we can solve \eqref{eq:min-max:baseline}, then we can guarantee that the new controller is an improvement over the baseline. While it may not be possible to satisfy these conditions in practice, we will in this paper discuss how we can satisfy them with high probability, and show in numerical examples the benefit of introducing a baseline controller.

An illustration of the difference between \eqref{eq:min-max} and \eqref{eq:min-max:baseline} using a simple P-controller $C(q) = K$ is given in Figure~\ref{fig:1D-case-set}. 
Here the red dashed line indicates the optimal controller with respect to the unknown system $G_\circ$, while the blue dashed line indicates the gain of the baseline controller $K_b$. Due to the uncertainty some systems in $\mathcal{G}$ have an optimal gain lower than the baseline controller, while other systems have an optimal gain higher than the baseline. The min-max strategy in \eqref{eq:min-max} then chooses a very conservative controller with a gain close to zero, but the safe improvement strategy \eqref{eq:min-max:baseline} stays at the baseline since it cannot be certain that any of the other possible controllers will improve the performance. 

In the remainder of this paper, we will discuss and evaluate how \eqref{eq:min-max:baseline} can be adapted to the setting of linear output feedback control. 
In Section~\ref{sec::sets} we propose one way of designing the sets $\mathcal{G}$ and $\mathcal{C}$. Unfortunately, even when the system is linear, the optimization problem \eqref{eq:min-max:baseline} is still hard, so in Section \ref{sec::convex} we use a similar strategy as \cite{anna2019bayesian} to find a convex approximation of this optimization problem. Finally, in Section~\ref{sec::num} we provide some numerical examples.

\section{The model uncertainty and Controller Set}\label{sec::sets}
To show how a baseline controller can be incorporated into existing data-driven output feedback schemes, we consider the model reference control problem as in \cite{anna2019bayesian}. That is, the control objective is to design a controller
such that the closed-loop system
\begin{align}
   \frac{G_{\circ}(q)C(q)}{1+G_{\circ}(q)C(q)}, \nonumber
\end{align}
is close to a desired reference model $W(q)$.
Here, we assume that $W(q)$ is given, but it can be designed by system identification methods or features of the desired system \cite{aastrom2021feedback}. 
To evaluate the closeness of the resulting closed-loop system to the reference model, a reasonable performance criterion used by e.g. \cite{anna2019bayesian, ferizbegovic2021bayes} is
 \begin{align}
 \label{eq:criterion}
  J(C, G_{\circ}) =  \left \Vert W(q)-\frac{G_{\circ}(q)C(q)}{1+G_{\circ}(q)C(q)} \right \Vert_2^2,
 \end{align}
To find an uncertainty set, we take a system identification approach. While different types of model structures can be considered, we here use Finite Impulse Response (FIR) models, which have the benefit of giving a simple way of determining the parameter uncertainties even with finite samples. By considering FIR models of high orders, we can capture a general class of stable linear dynamic systems, but high-order models also lead to higher uncertainty in the estimates. We will see in the numerical examples that in such a setting the use of a baseline controller can improve performance while it at the same time reduces the variance of the controller parameters. 

The FIR model is defined as
\begin{align}
   G(q) = \sum_{i=0}^{n-1} g_i q^{-i}
\end{align}
where $g = (g_i)_{i=0}^{n-1}$ is the unknown impulse response. For the theoretical development, we here assume that the true system $G_o$ can be represented as an FIR model with large enough $n$. However, in the numerical examples, we will consider systems with infinite impulse response and approximate them with FIR models of high. 
For more compact notation, let
\begin{align}
\begin{aligned}
\varphi_t=\begin{bmatrix}u_{t}&\dots &u_{t-n+1}\end{bmatrix}^T, \\
\Phi = \begin{bmatrix}\varphi_0&\dots &\varphi_{N-1}\end{bmatrix}^T,\\
Y=\begin{bmatrix}y_0&\dots &y_{N-1}\end{bmatrix}^T.    
\end{aligned}
\end{align}
We assume that the input is exciting enough to give a $\Phi$ with full column rank and that $v_t \sim \mathcal{N}(0, \sigma^2_v)$ is a white noise process. Then a least squares estimation of $g$ gives 
\begin{align}\label{eq:ghat}
\begin{aligned}
    \hat{g}&=(\Phi^T\Phi)^{-1}\Phi^TY,\\
	\Sigma &=  \sigma_v^2(\Phi^T\Phi)^{-1}
\end{aligned}
\end{align} where $\hat{g}$ denotes the estimation of $g$ and $\Sigma$ is the covariance matrix.
Since $\sigma_v^2$ is unknown, an unbiased estimate can be used
\begin{align}
    \hat{\sigma}_v^2 = \frac{1}{N-n-1}
    \left \Vert 
    (Y-\Phi\hat{g}) \right \Vert_2^2
\end{align} 
Consequently, the estimated covariance is given by $\hat{\Sigma}=\hat{\sigma}_v^2(\Phi^T\Phi)^{-1}$. 
An uncertainty set can finally be obtained by
\begin{align}\label{eq:Gset}
    \mathcal{G}& =  \{
    g: (g-\hat{g})^T\hat{\Sigma}^{-1}(g-\hat{g}) \leq q_{\alpha}
    \}
\end{align}
where $q_{\alpha}$ denotes the $\alpha$-quantile of distribution $F(n,N-n)$.
Then, the true impulse response $g_\circ\in \mathcal{G}$ with a probability no smaller than $1-\alpha$ asymptotically \cite{rao1973linear-statstics}.

For the controller set $\mathcal{C}$, we simply assume that it is given as a linear combination of $p$ basis operators $\phi_i(q)$, i.e. that
\begin{align}
 	\mathcal{C} = \left\{C_\rho\,:\, C_\rho = \sum_{i=1}^p \rho_i \phi_i(q) \right\}.
  \label{eq:linear_controller}
 \end{align}
An example of basis operators is shown in Section~\ref{sec:num::high}.
\section{Convex approximation}\label{sec::convex}
The criterion in \eqref{eq:min-max:baseline} is not convex-concave and is in general hard to solve. The main issues are that
 \begin{itemize}
 	\item $J(C,G)$ is not convex in $C$,
    \item and the set $\mathcal{G}$ is uncountable.
 \end{itemize}
To overcome this we use two approximations, first a convex approximation of $J(C,G)$ is presented in Section~\ref{sec:convexJ} and then the scenario approach is used in Section~\ref{sec:scenario} to get a finite set $\mathcal{G}_M$.

\subsection{Approximation of the criterion}\label{sec:convexJ}
In \cite{anna2019bayesian, formentin2018core}, it is proposed that with a good controller $C(q)$ we get
\begin{align}
	W(q) \simeq  \frac{G(q)C(q)}{1+G(q)C(q)} \nonumber
\end{align}
and thus
\begin{align}
	1-W(q) \simeq  \frac{1}{1+G(q)C(q)} \nonumber .
\end{align}
Then, an approximated performance criterion based on \eqref{eq:criterion} can be defined as
\begin{align}
\label{eq:criterion:tildeJ-baseline}
\begin{aligned}
     J(C,G) &= \left \Vert W(q)-\frac{G(q)C(q)}{1+G(q)C(q)} \right \Vert_2^2 \\
 &\simeq
	\Vert 
 W(q)-\left(1-W(q)\right)G(q)C(q)
	\Vert_2^2 \\
 &:= \tilde{J}(C,G)
\end{aligned}
\end{align}
Replacing $J(C, G)$ with $\tilde{J}(C,G)$ in \eqref{eq:min-max:baseline} we approximate the baseline regret according to

 \begin{align}
    \label{eq:minmax:tildeJ-baseline}
 	 \min_\rho \max _{G\in \mathcal{G}}  \tilde{J}(C_\rho, G) - \tilde{J}(C_b,G)
 \end{align}
Note that this criterion in combination with linear-in-parameters controllers \eqref{eq:linear_controller} is convex in $\rho$. Furthermore, the optimization problem \eqref{eq:minmax:tildeJ-baseline} can be rephrased as~\cite{boyd2004convex},
\begin{align}
\label{eq:sa}
\begin{aligned}
& \min_{\rho,\beta}~\beta \\
& \text{s.t.}~
\tilde{J}(C_\rho,G) 
- \tilde{J}(C_b,G)	
\leq \beta,~
\forall G \in \mathcal{G}
\end{aligned}
\end{align}

\subsection{The scenario approach}\label{sec:scenario}
The optimization problem in \eqref{eq:sa} still has an uncountable set of constraints.
To approximate it with a finite number of constraints the scenario approach \cite{calafiore2006scenario, campi2021scenario} is considered. That is, a finite number of $M$ samples are drawn from $\mathcal{G}$ to create a finite set $\mathcal{G}_M \subset \mathcal{G}$. 
Here we use a truncated normal distribution so that we draw each $g_i$ from the distribution $\mathcal{N}(\hat{g}, \hat{\Sigma})$, see \eqref{eq:ghat}, but reject the sample if it is not in $\mathcal{G}$.

This finally gives us the convex optimization problem
\begin{align}
\label{eq:criterion:sa}
\begin{aligned}
&\min_{\rho, \beta}~\beta \\
&\text{s.t.}~
\tilde{J}(C_\rho,G_i) 
- \tilde{J}(C_b,G_i)	
\leq \beta,~
\forall G_i \in \mathcal{G}_{M}.
\end{aligned}
\end{align}
Here the value of $M$ guarantees a certain level of robustness with desired confidence. It can also be seen as a hyperparameter to choose the balance between computation and the desired performance. 

By noticing that the optimal solution to \eqref{eq:criterion:sa} always has $\beta \leq 0$ we can restate \cite[Theorem 1]{campi2009scenario} as follows.
\begin{theorem}
    Fix two real number, the level parameter $\epsilon \in (0,1)$  and confidence parameter $\eta \in (0,1)$. If 
    \begin{align}
		M \geq \frac{2}{\epsilon} \left (\ln \frac{1}{\eta}+p  \right ), 
    \label{eq:theorem:M}
	\end{align} 
	(recall that $p$ is the number of controller parameters), then with a probability no smaller than $1-\eta$,  
    the resulted controller $\hat{C}^*$ of solution $\hat{\rho}^*$ is no worse than the baseline controller $C_b$ for all $G \in \mathcal{G}$ but at most $\epsilon$-fraction, i.e. $\mathbb{P}
    \left(\tilde{J}(C_\rho,G_i) - \tilde{J}(C_b,G_i) \nleq 0 \right)\leq \epsilon $.

\end{theorem}

\begin{figure*}[htbp]
   \centering
   \subfloat[$N=200$]{
    \includegraphics[width=0.9\textwidth]{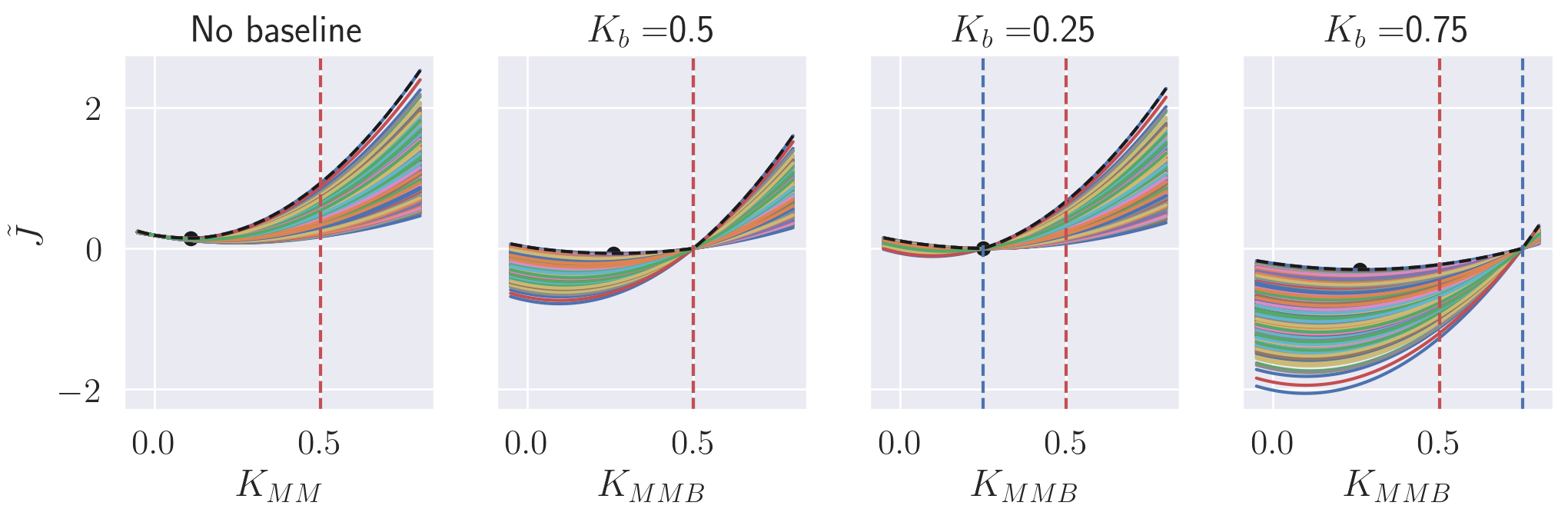}
   \label{fig::1D-200}
   }
   \\
   \centering
   \subfloat[$N=1000$]{
   \includegraphics[width=0.9\textwidth]{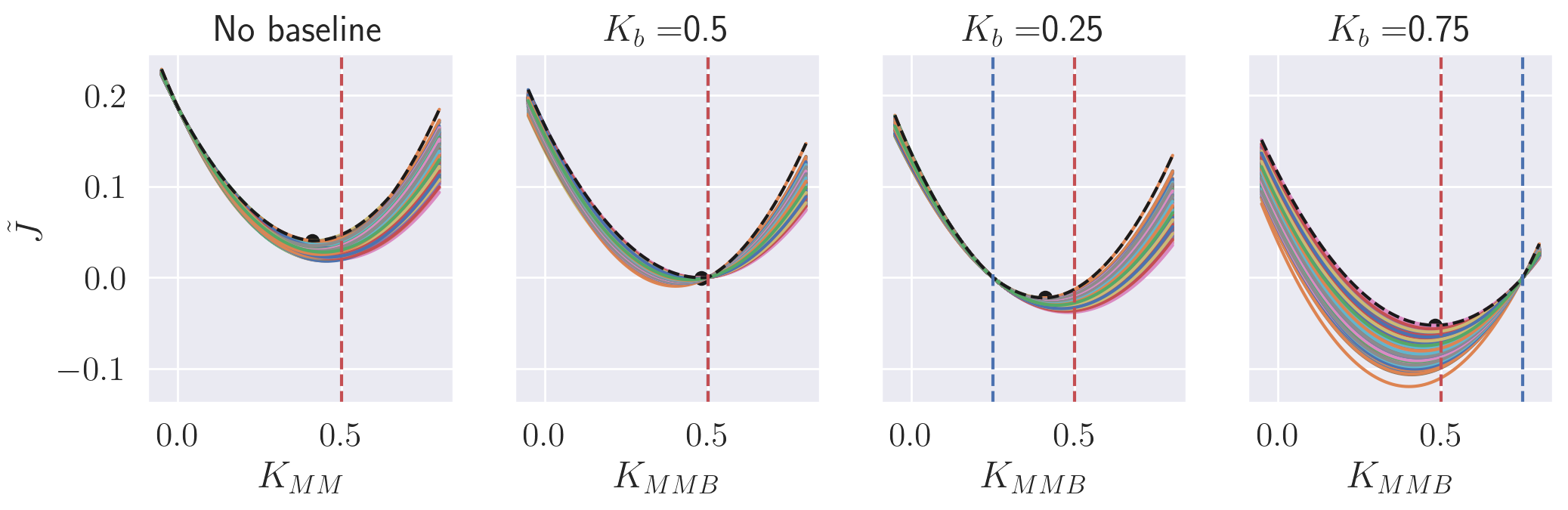}
   \label{fig::1D-1000}
   }
   \\
\caption{Experiment comparison between the proposed method \textbf{without and with different baseline controllers}.
The optimization problem is solved with {$M=1523$} sampled scenarios from $\mathcal{G}$ represented by the \emph{colored quadratic lines}.
The \emph{black dashed line} denotes the worst-case cost over these scenarios and the dot is the solution. The \emph{red dashed line} indicates the optimal controller and
the \emph{blue dashed line} indicates the baseline controllers.
{
The results demonstrate that incorporating a baseline \textbf{reduces conservatism} and results in a controller gain \textbf{closer to the optimal value}, particularly with limited data.
}
}

\label{fig::1D}
\end{figure*}

\section{NUMERICAL EXAMPLE}\label{sec::num}
Two numerical examples are studied to evaluate the methods discussed in this paper.
For the experiments $\alpha=0.01$ was used to create the uncertainty set in \eqref{eq:Gset}, and the number $M$ of sampled systems in $\mathcal{G}_M$ was chosen as the lower bound of \eqref{eq:theorem:M} with $\epsilon=0.01$ and $\eta = 0.05$. 
{
Specifically, $M=1523$ was computed for the 1D P-controller tuning example in Section V.A and $M=6138$ for the high-dimensional case in Section V.B. 
}

\subsection{Tuning a P-controller}\label{sec::num::1d}
To illustrate the difference between the pure min-max criterion and the safe controller improvement  we first consider a simple example where a proportional controller $C(q) = K$
is tuned. 
The true system is a simple second-order Butterworth filter,
\begin{align}
   G_{\circ}(q) = \frac{0.1004q^2+0.2008+0.1004 }{q^2 - 1.561q + 0.6414}
\end{align}
The reference model is chosen as
\begin{align*}
   W(q)&=\frac{K_o G_{\circ}(q)}{1+K_o G_{\circ}(q)}\\ 
   &=\frac{0.00994q^2+0.01989q+0.00994}{q^2 - 1.526 q + 0.645}
\end{align*} 
where $K_o = 0.5$. This means that the criterion \eqref{eq:criterion} is optimized for $K=0.5$.

To construct $\mathcal{G}$ in \eqref{eq:Gset},
open-loop data was collected using a white noise input with a variance $1$. The measurements were corrupted by white Gaussian noise with a variance of 0.5. 
An FIR model of order $n=100$ was used in the estimation.

The results presented in Figure~\ref{fig::1D} demonstrate how the scenario approach selects its optimized controller and the influence of different baseline controllers. Here $K_b$ denotes the gain of the baseline controller, $K_{\text{MM}}$ refers to the gain of the min-max controller that optimizes $\tilde{J}(C,G)$ and $K_{\text{MMB}}$ is the controller obtained by \eqref{eq:criterion:sa}.

As shown in Figure~\ref{fig::1D-200}, with only $200$ input-output pairs, the estimated system is highly uncertain, as evidenced by the spread of the different scenarios. 

In this case, the optimized P-controller without a baseline controller $K_{\text{MM}}$ has a gain close to zero, suggesting that no action should be taken. 
This is not surprising because with a limited number of samples and with the observation noise, the parametric uncertainty of the model is high. 

This situation also affects the solution $K_{\text{MMB}}$. 
When the baseline controller is $K_b=0.5$, i.e. the optimal controller, the solution is still not optimal. The reason for this is that even if $\mathcal{G}$ contains the true system, the scenario approach employed here only uses a finite number of constraints. 
However, even in this case, we get a less conservative controller than the pure min-max approach. 

When we look at the two more realistic scenarios where the baseline gain is either higher or lower than the optimal gain, then we see that \eqref{eq:sa} finds a controller gain that is either the same as, or an improvement over, the baseline gain. 

With larger data size, as depicted in
Figure~\ref{fig::1D-1000},  a clear increase in the density of lines representing sampled scenarios is observed, with the gain of all estimated controllers converging towards the optimal gain $0.5$.
That is, as the uncertainty set $\mathcal{G}$ shrinks, $K_{\text{MM}}$ and $K_{\text{MMB}}$ move closer to the optimal controller.

\subsection{High-Dimensional Case}\label{sec:num::high}
We consider the following 4th-order linear system, also used in \cite{pillonetto2010system_source, anna2019bayesian},
\begin{align}
	G_{\circ}(q)=\frac{0.28261q+0.50666}{
	A(q)}
\end{align} where $A(q)=q^4-1.41833q^3+1.58939q^2-1.31608q+0.88642$.
For the estimated FIR model, the order is set to $n=200$, and the selected controller class is expressed as
{
\begin{align}
	C_\rho(q) = \frac{\sum_{k=0}^{5}\rho_k q^{-k}}{1-q^{-1}},
\end{align}
}which combines an integrator with a FIR model of order~$5$.
The desired closed-loop system is formulated as
\begin{align}
	W(q)=\frac{C_{\rho^*}(q)G_{\circ}(q)}{1+C_{\rho^*}(q)G_{\circ}(q)},
\end{align}
with 
$$\rho^*=\begin{bmatrix} 0.20& -0.27& 0.29& -0.24& 0.16& 0.0084\end{bmatrix}^T.$$
The baseline controller is set to
$$\rho_b=\begin{bmatrix}0.06 &  0.02 & -0.03 & 0.00  &  0.03 &  0.02\end{bmatrix}^T.$$
Denote $\rho_{\text{MMB}}$ as the controller parameters for the solution to \eqref{eq:criterion:sa} and $\rho_{\text{MM}}$ be that without baseline controller.
To assess the advantage of considering the min-max criterion when the estimation is uncertain, we will also compare the performance with 
 \begin{align}
     \rho_{\text{nom}} = \min_\rho \tilde{J}(C_\rho, \hat{G})
     \label{eq:rho_nom}
 \end{align}
 where $\hat{G}$ is the mean estimated FIR-model found in \eqref{eq:ghat}, and thus does not take uncertainty into account. 

Similarly to in Section~\ref{sec::num::1d}, the data for system identification is collected using white inputs with $N=300, 800, 1300$ respectively, and outputs are then corrupted by white Gaussian noise 
{with a signal-to-noise-ratio (SNR) of 10. }

\begin{figure}
   \centering
   \subfloat[$F_W$ with $N=300$]{   
   \includegraphics[width=0.49\linewidth]{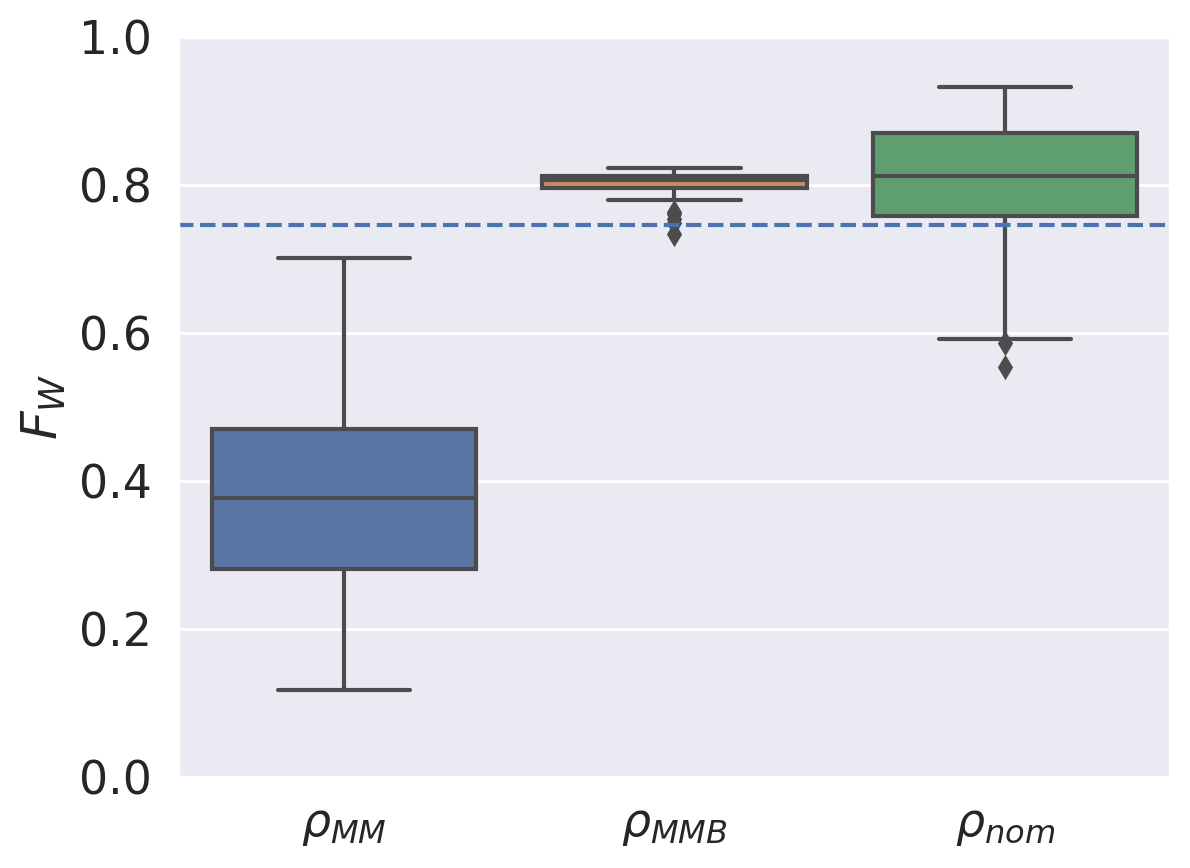}
   \label{fig::HDFW-300}
   }
   \centering
   \subfloat[$F_C$ with $N=300$]{   
   \includegraphics[width=0.49\linewidth]{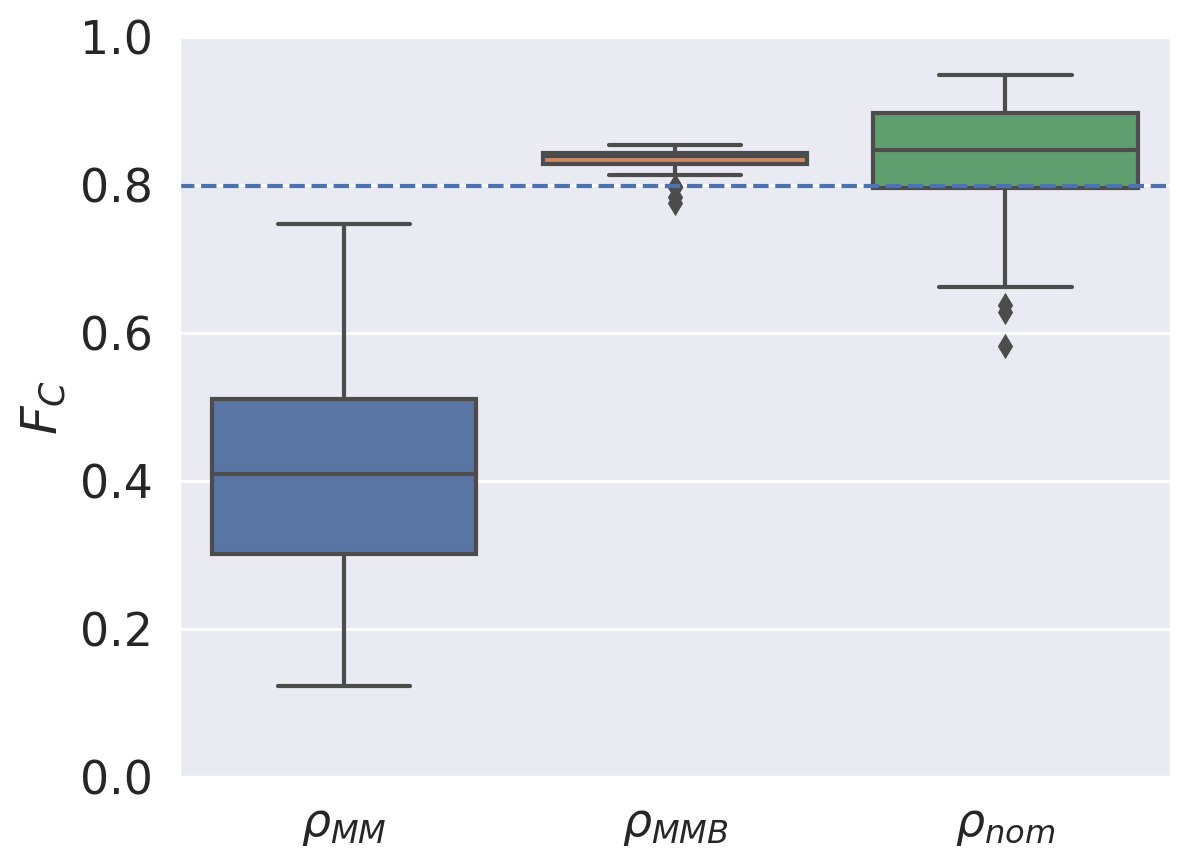}
   \label{fig::HDFC-300}
   }
   \\
   
   \subfloat[$F_W$ with $N=800$]{
   \includegraphics[width=0.49\linewidth]{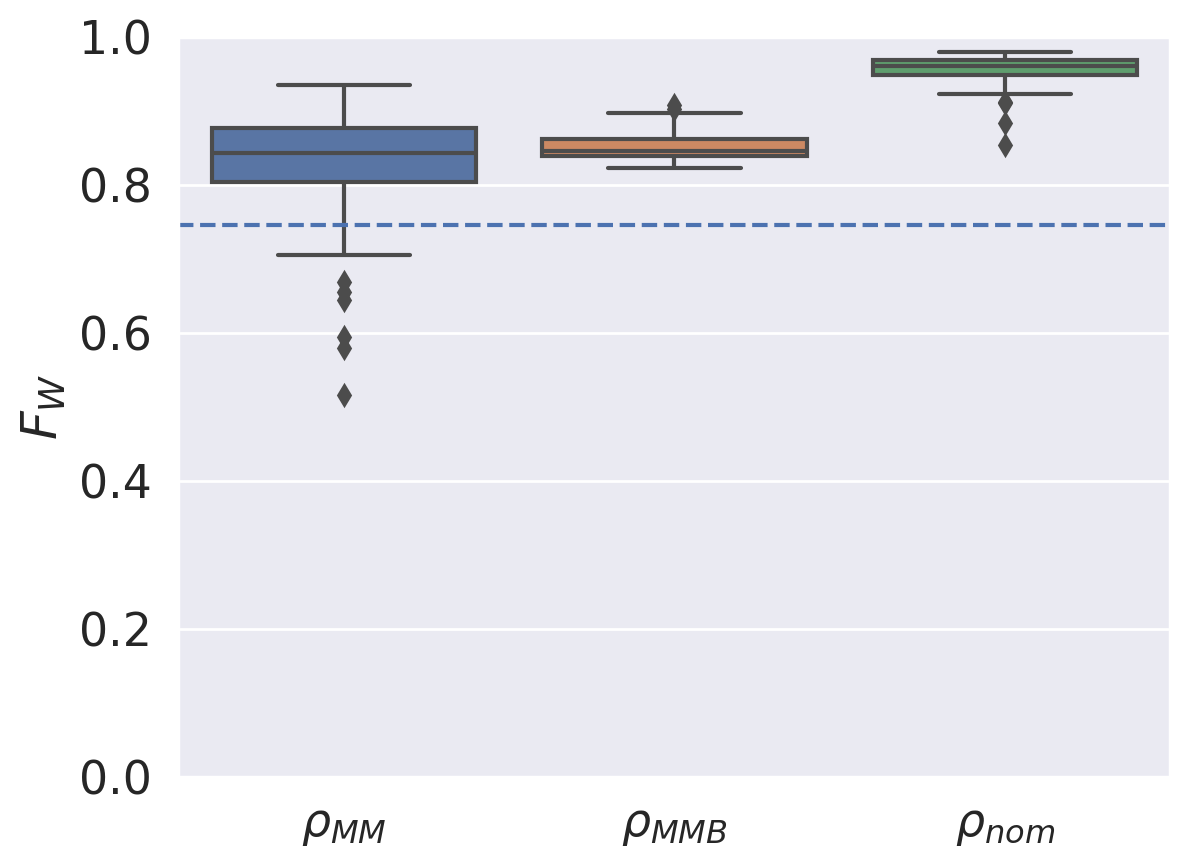}
   \label{fig::HDFW-800}
   }
   \subfloat[$F_C$ with $N=800$]{
   \includegraphics[width=0.49\linewidth]{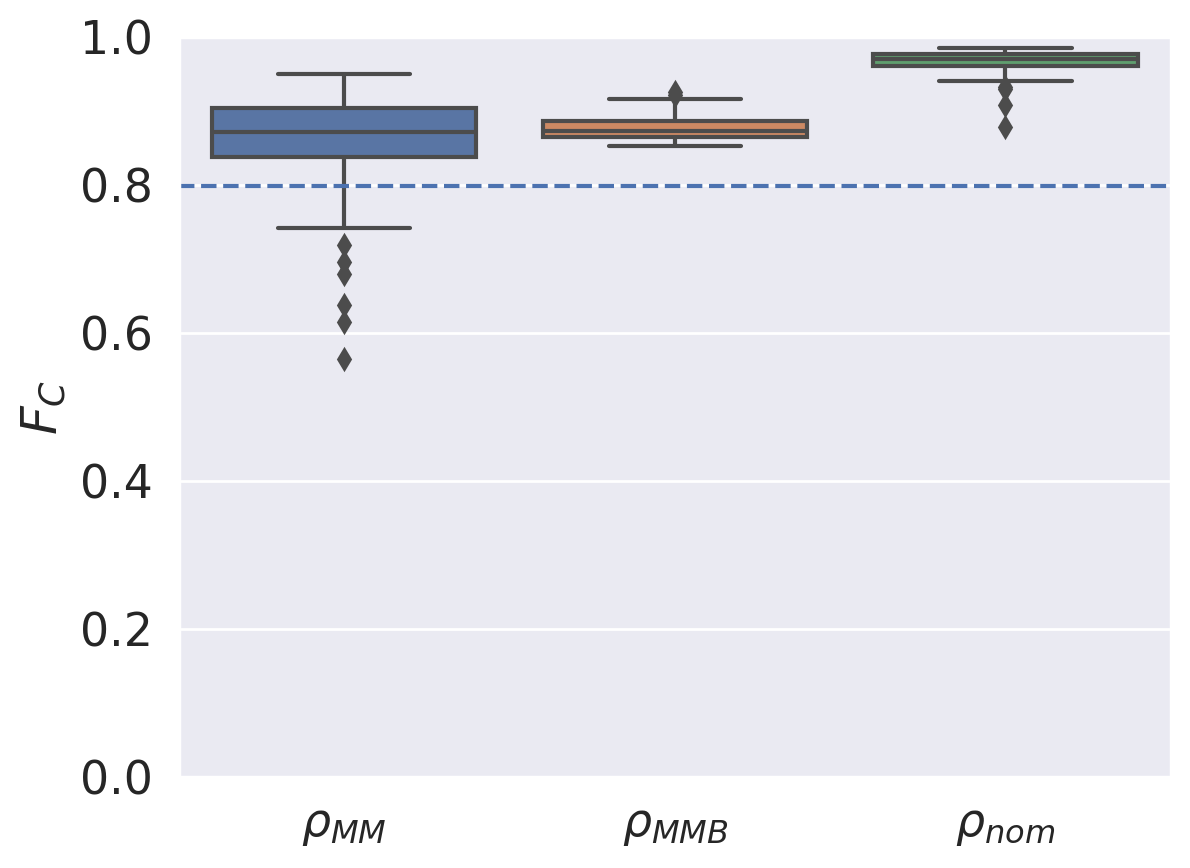}
   \label{fig::HDFC-800}
   }
   \\
   
   \subfloat[$F_W$ with $N=1300$]{
   \includegraphics[width=0.49\linewidth]{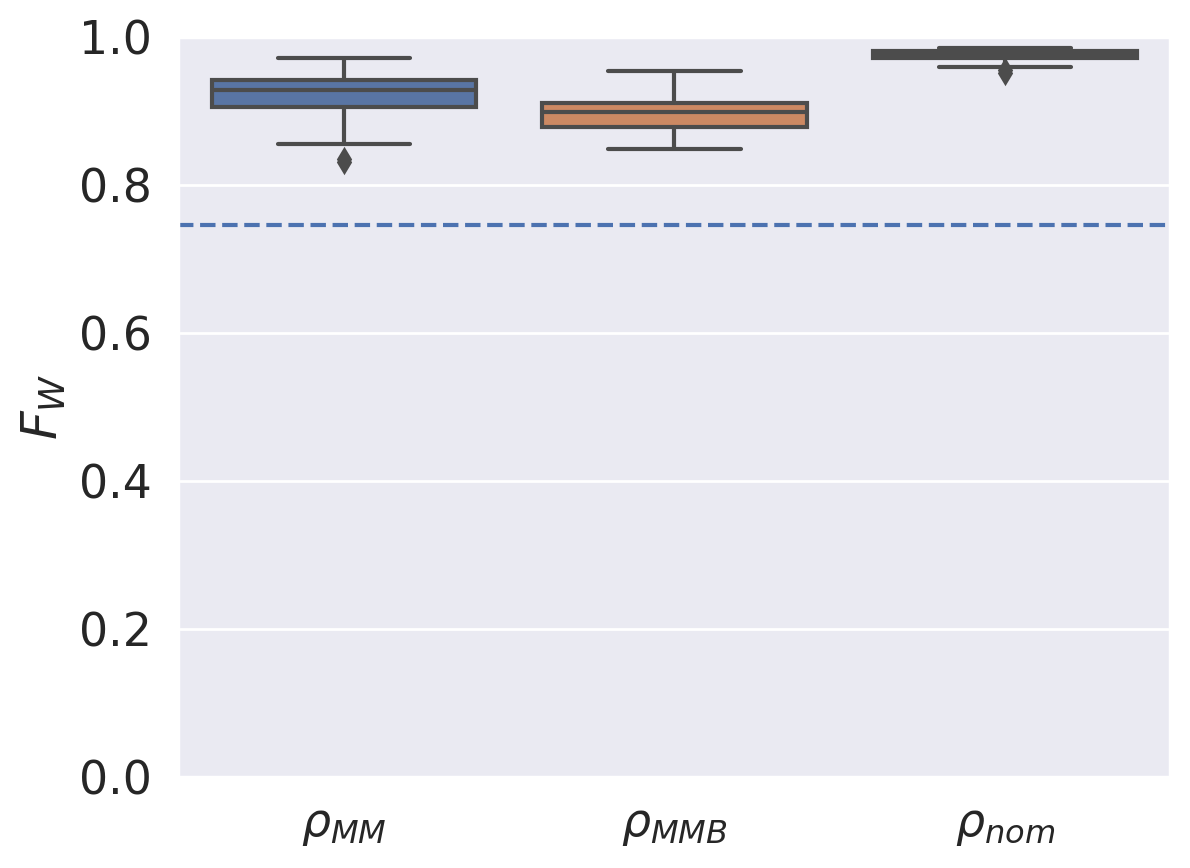}
   \label{fig::HDFW-1300}
   }
   \subfloat[$F_C$ with $N=1300$]{
   \includegraphics[width=0.49\linewidth]{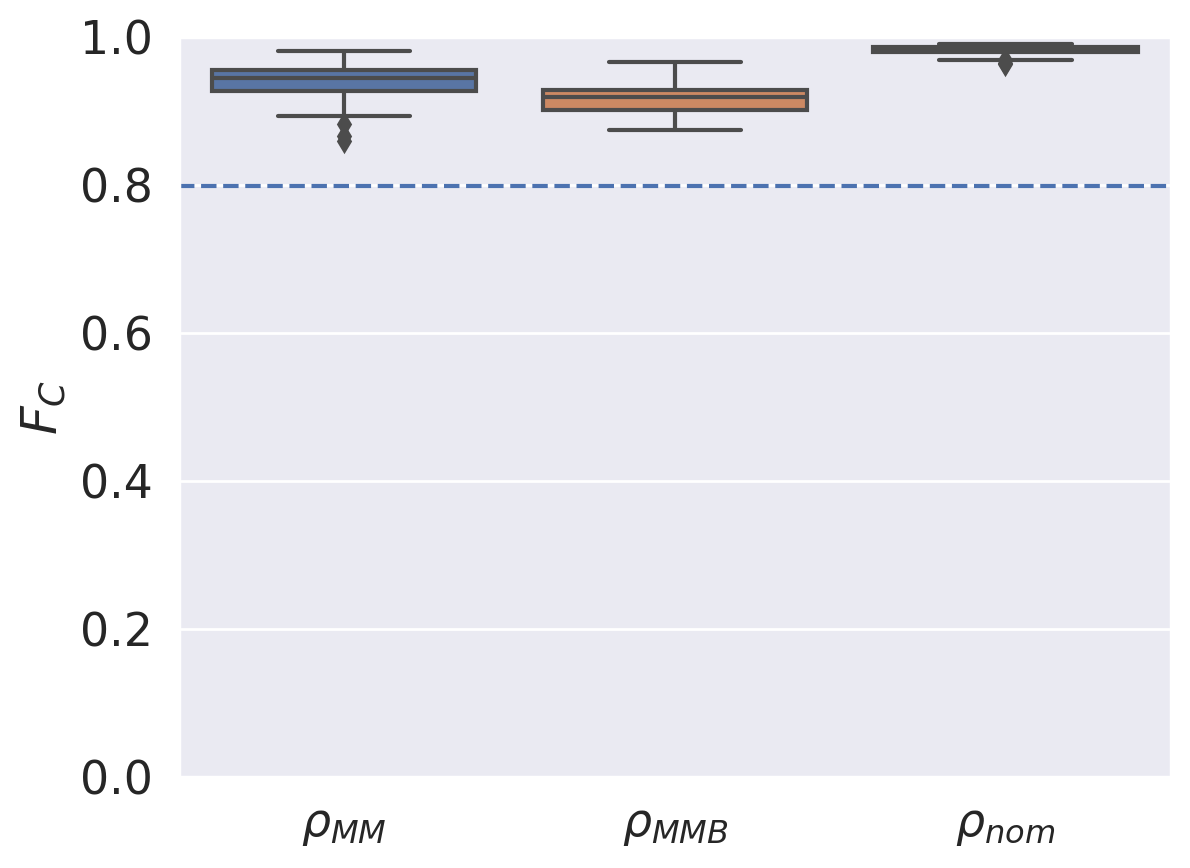}
   \label{fig::HDFC-1300}
   }

 \caption{Performance comparison under 100 runs w.r.t $F_W$ in \eqref{eq:FW} and $F_C$ in \eqref{eq:FC} between $\rho_{\text{nom}}$ in \eqref{eq:rho_nom}, $\rho_{\text{MMB}}$ (min-max \textbf{with baseline}) and $\rho_{\text{MM}}$ (min-max \textbf{without baseline} controller) with the \textbf{different number of samples} to identify the model parameters. 
 {The optimization problem is solved with $M=6138$ sampled scenarios.}
 The \emph{blue dash line} is the performance of the baseline controller. The \emph{box} shows the quantiles of the runs, while the \emph{whiskers} extend to the 99.3\% coverage and the black dots represent the outliers.
 }
 \label{fig::HD}
 \end{figure}

To evaluate the accuracy of the results,  two metrics are used. The first, $F_W$, quantifies the fitting error between the reference model and the actual closed-loop system,
\begin{align}
F_W=1-
\frac{
 	J(C_\rho, G_{\circ})
 	}{\Vert W\Vert_2^2}.
\label{eq:FW}
\end{align} 
The second evaluation metric $F_C$ relates to \eqref{eq:criterion:tildeJ-baseline}, defined as
 \begin{align}
 	F_C=1- \frac{
 	\tilde{J}(C_\rho, G_{\circ})
 	}{\Vert W\Vert_2^2}.
\label{eq:FC}
 \end{align}
We perform the experiment with a Monte Carlo analysis of $100$ tests. For each run, the number of sampled systems in the scenario approach is set to $M= 6138 $ to guarantee the robustness and confidence of the optimization result with $\epsilon=0.01$ and $\eta=0.05$.
The results shown in Figure~\ref{fig::HD} illustrate the distribution of the $100$ runs.

As shown in Figure~\ref{fig::HDFW-300} and Figure~\ref{fig::HDFC-300}, with a limited number of data, the estimated controller $\rho_{\text{MM}}$ without any baseline does not work well. 
The mean value of bot fir terms $F_W$ and $F_C$ is around 0.4 indicating the resulting closed-loop system is quite far away from the reference model.
As for the nominal controller $\rho_{\text{nom}}$, the mean performance is slightly above that of the baseline controller but the performance has a large variance. For some extreme runs, the fit from $\rho_{\text{nom}}$ is around $0.55$. 
On the contrary, when the baseline is added the resulting design is robust without leading to a conservative controller. 
The \emph{whiskers} in the box plot of $\rho_{\text{MMB}}$, representing 99.3\% coverage, consistently surpasses the baseline controller.
This indicates that our method can guarantee the estimated controller $\rho_{\text{MMB}}$ is at least as good as $\rho_b$ with high probability.

As the number of collected data increases, the uncertainty of the estimated model diminishes, as illustrated in the second row of Figure~\ref{fig::HD}. 
Here, the mean performance of the $\rho_{\text{MM}}$ is similar to that of $\rho_{\text{MMB}}$, though the variance in $\rho_{\text{MM}}$ performance is markedly higher. 
The nominal controller $\rho_{\text{nom}}$ works well in this case but it does not consider the uncertainty. In real-world applications, it is challenging to ascertain when sufficient data has been acquired to estimate a viable controller. 
When collecting $1300$ input-output pairs, all three controllers outperform the baseline controller $\rho_b$ (\emph{blue dashed line}).

\section{CONCLUSIONS}\label{sec::conclusions}
This paper addresses  \emph{safe} improvement of a given baseline controller for an unknown system. This is achieved by augmenting a pure min-max strategy with a baseline regret.
The methodology encompasses two stages: uncertainty set determination and controller optimization. 
With the uncertainty associated with the estimated system, the controller is refined by solving convex optimization problems.
In two numerical examples, it is shown that the addition of a baseline regret term to the min-max criterion leads to improved performance and decreased variance for the resulting controller. 

Since the scenario approach can not guarantee the performance on the unknown plant $G_\circ$ with a finite number of constraints, future works include the direct optimization of  \eqref{eq:minmax:tildeJ-baseline} or approximate \eqref{eq:min-max:baseline}. 
{
Also, while the current work focuses on SISO systems and model reference control, using baseline regret instead of a pure min-max strategy can be applied more generally. For example, it would be of interest to extend this idea to MIMO systems, using different system identification methods and performance criteria.
}
\newpage
\bibliographystyle{./IEEEtran} 
\bibliography{./IEEEabrv, ./mms}


\end{document}